\documentclass[bibyear]{aa}
\usepackage{graphicx}
\usepackage[varg]{txfonts}
\usepackage{natbib}
\begin{document}
\title{Flux calibration of medium-resolution spectra from 300\,nm to
  2\,500\,nm: Model reference spectra and telluric
  correction\thanks{The reference model spectra described here are
    available in electronic form at the CDS via anonymous ftp to
    cdsarc.u-strasbg.fr (130.79.128.5) or via http://cdsweb.u-strasbg.fr/cgi-bin/qcat?J/A+A/568/A9}
}
\titlerunning{Flux calibration of medium-resolution spectra from 300\,nm to
  2\,500\,nm}
\author{S.\,Moehler\inst{1} 
\and A.\,Modigliani\inst{1} 
\and W.\,Freudling\inst{1}
\and N.\,Giammichele\inst{2} 
\and A.\,Gianninas\inst{3} 
\and A.\,Gonneau\inst{4} 
\and W.\,Kausch\inst{5} 
\and A.\,Lan\c{c}on\inst{4} 
\and S.\,Noll\inst{5} 
\and T.\,Rauch\inst{6} 
\and J.\,Vinther\inst{1} } 
\institute{European Southern Observatory, Karl-Schwarzschild-Str. 2, D
  85748 Garching, Germany, 
\email{[smoehler;amodigli;wfreudli;jvinther]@eso.org}
\and D\'epartement de Physique, Universit\'e de Montr\'eal, CP. 6128,
Succursale Centre-Ville, Montr\'eal, QC H3C 3J7, Canada,
\email{noemi@astro.umontreal.ca}
\and Homer L. Dodge Department of Physics and Astronomy, University of
Oklahoma, 440 W. Brooks St., Norman, OK, 73019, USA, \email{alexg@nhn.ou.edu}
\and Observatoire astronomique de Strasbourg, Universit\'e de Strasbourg, CNRS,
UMR 7550, 11 rue de l'Universit\'e, F-67000 Strasbourg, France
\email{[anais.gonneau;ariane.lancon]@astro.unistra.fr} 
\and Institut f\"ur Astro- und Teilchenphysik, Universit\"at Innsbruck,
Technikerstr. 25/8, 6020 Innsbruck, Austria
\email{[Wolfgang.Kausch;Stefan.Noll]@uibk.ac.at}
\and Institute for Astronomy and Astrophysics, Kepler Center for Astro
and Particle Physics, Eberhard Karls University, Sand 1, 72076 T\"ubingen,
  Germany (\email{rauch@astro.uni-tuebingen.de})
} \date{Received 10 March 2014 / Accepted 7 April 2014 }
\abstract
{While the near-infrared wavelength regime is becoming more and more
  important for astrophysics there is a marked lack of
    spectrophotometric standard star data that would allow the flux
  calibration of such data. Furthermore, flux calibrating medium- to high-resolution \'echelle spectroscopy data is challenging even in the
  optical wavelength range, because the available flux standard data
  are often too coarsely sampled.}
{We will provide standard star reference data that allow users to
  derive response curves from 300\,nm to 2\,500\,nm for 
  spectroscopic data of medium to high resolution, including those taken
  with \'echelle spectrographs. In addition we describe a method to correct for
  moderate telluric absorption without the need of observing telluric
  standard stars.}
{As reference data for the flux standard stars we use theoretical spectra
  derived from stellar model atmospheres. We verify that they provide
  an appropriate description of the observed standard star spectra by
  checking for residuals in line cores and line overlap regions in the
  ratios of observed (X-shooter) spectra to model spectra. The finally
  selected model spectra are then corrected for remaining mismatches
  and photometrically calibrated using independent observations. The
  correction of telluric absorption is performed with the help of
  telluric model spectra. }
{We provide new, finely sampled reference spectra without telluric
  absorption for six southern flux standard stars that allow the users
  to flux calibrate their data from 300\,nm to 2\,500\,nm, and a
  method to correct for telluric absorption using atmospheric models.}
{}
\maketitle

\section{Introduction}\label{sec:intro}

Accurate flux calibration of astronomical spectra remains a
significant challenge. Spectral flux calibration requires flux
calibrators with known absolute fluxes that are accessible with the
same spectrograph that also takes the spectra of the science
targets. With the arrival of new generations of spectrographs that
cover wide wavelength ranges and produce relatively high-resolution
spectra, known and well-tested spectrophotometric standard star
catalogues as listed by \citet{oke90} and \citet{hamuy92,hamuy94} are
no longer adequate for spectral flux calibration because they do not
extend to the near-infrared (NIR) and/or are  too coarsely
sampled to permit the flux calibration of high-resolution spectra. For
example, the European Southern Observatory's (ESO) X-shooter
instrument covers, in a single exposure, the spectral range from
300\,nm to 2500\,nm and operates at intermediate spectral resolution
($R$$\approx$4000--17\,000, depending on wavelength and slit width)
with fixed \'echelle spectral format in three optimized arms (UVB:
300\,nm--550\,nm, VIS: 550\,nm--1000\,nm, NIR: 1000\,nm--2500\,nm; see
\citealt{vede11} for more details).

In this paper we present a new set of calibrated model spectra for
flux standard stars covering the wavelength range from 300\,nm to
2\,500\,nm. These spectra are useful for deriving consistent instrumental
response curves over this wide range of wavelengths with a spectral
resolving power of up to 40\,000 and possibly more.  We also describe in
detail how to use the spectra in this manner.

Our approach to obtain fully calibrated model spectra was as
follows. First, we selected a set of flux standard stars observable
from the southern hemisphere that cover the full right ascension
range, whose spectra can be modelled accurately, and for which
X-shooter spectra exist (Sects.\,\ref{ssec:sample}, \ref{ssec:data},
and \ref{ssec:model_spectra}).  
We then used X-shooter observations of these stars to compute the
ratio of the observed spectra to the model spectra, i.e. the
instrumental response (see Sect.\,\ref{ssec:method}).  If the model
spectra perfectly described the spectra of all stars, response curves
derived from different stars observed with the same instrumental setup
should only differ by signatures imposed by the atmosphere (e.g.
  telluric absorption, varying atmospheric transmission). However,
due to small deficiencies in the model spectra, for most of the stars
such ratios show star-specific features in regions dominated by
overlapping lines (see Fig.\,\ref{fig:Montreal}). To be able to
  identify and fit such deficiencies we observed a star that does not
  show any lines within the X-shooter spectral range and whose
  spectrum can be modelled accurately. We used the
  ratio between model and observations of that star together with the
observed spectra of all our stars to derive corrections to their
model spectra (Sect.\,\ref{ssec:bump_UVB}). Finally, we use the
available (spectro-)photometric data for the stars to compute the
absolute flux scale of the corrected model spectra (see
Sect.\,\ref{ssec:phot_cal}).  The result is a self-consistent new set
of fully flux calibrated spectrophotometric standards for the southern
hemisphere.

In order to compute response curves for an instrument from observed
spectra and the models, the effect of the Earth's atmosphere must be
removed from the observed spectra.  In Sect.\,\ref{sec:atmos},
we describe a method for such a correction that we used for X-shooter
spectra. This includes a fast and efficient removal of the telluric
absorption feature, that is sufficiently accurate for the intended
purpose of deriving response curves and can be adapted for any other
instrument providing spectra of sufficient resolution between
  600\,nm and 2\,500\,nm.  Finally, in Sect.\,\ref{sec:pipe}, we
describe in detail the full procedure of our approach to compute
response curves for the X-shooter instrument, i.e.  medium-resolution
\'echelle spectra covering a very wide wavelength range.

\section{Reference spectra}

\subsection{Sample selection}\label{ssec:sample}

Spectral flux calibration utilizes as reference either well-calibrated observations or a spectral model of a standard star.  The
advantage of using a model is that it is noiseless, and does not
include features imposed by the terrestrial atmosphere.  The process
of flux calibration requires computing the ratio of an observed
spectrum with the model, and such a ratio is less sensitive to errors
in the wavelength scale if the spectrum is smooth and featureless.
Therefore, the ideal star to be used as a spectral flux standard has a
smooth and featureless spectrum that can be accurately modelled with a
minimum number of parameters.  Since spectral models in most cases
cannot predict the absolute scaling of the spectrum, the model must be
accompanied by accurate absolutely calibrated (spectro-)photometric
observations at some wavelengths within the wavelength range covered
by the spectrum.  Unfortunately, only few available standards satisfy
all criteria simultaneously. However, it should be noted that the
model of a star can be used to derive the {\it shape} of the response
curve of an instrument even if no absolute spectral calibration is
available. Such a spectrum can therefore be used to test and improve
the model spectra of other stars.

For the current work, we searched for stars with the following
criteria. In order to be able to model the spectra, we limited our
search to hot white dwarfs and hot subdwarfs. As a first step, we
limited our search to stars in the southern hemisphere, and to stars
with available flux information from \citet{hamuy92,hamuy94} or
the Hubble Space
Telescope\footnote{\url{http://www.stsci.edu/hst/observatory/crds/calspec.html/}}
and X-shooter spectra in the ESO archive. This selection resulted in
six standard stars, namely EG\,274, GD\,71, GD\,153, LTT\,3218,
LTT\,7987 (all hot DA white dwarfs), and Feige\,110 (a hot
subdwarf). We then extended our search to include at least one star in
the southern hemisphere that has a spectrum free of absorption lines
between 300\,nm and 2\,500\,nm, regardless of whether X-shooter
observations existed, and found \object{L97-3} (a white dwarf with a
featureless spectrum in the X-shooter wavelength range).

\subsection{Data}\label{ssec:data}

The six standard stars 
are routinely observed with X-shooter as part of its regular
calibration plan. We selected X-shooter spectra of these flux standard
stars observed in NODDING\footnote{This observation mode permits a
  good sky correction because it involves a series of short exposures
  with the object at different positions along the slit.} mode between
June 1, 2011 and July 5, 2012 (always referring to the beginning of the
night), which resulted in 203 spectra per arm, observed on 185
nights. We used data from nights of any photometric quality as we are
interested in the shape of the response curves and not in their
absolute level.

Figure\,\ref{fig:LTT7987_obs} shows example X-shooter spectra of the
flux standard star LTT\,7987, that clearly show the deep and wide
stellar lines (marked in red) in the blue (top), which may extend
across more than one \'echelle order, and the telluric absorption
(marked in green) in the redder part of the wavelength range (middle
and bottom).

The DC white dwarf L97-3 was observed on the nights October 1,
2012, October 20, 2012, and December 21, 2012, in the same way as the
flux standard stars listed above, i.e. with a slit width of
5\arcsec\ and in NODDING mode, with total exposure times between
1200\,sec and 1360\,sec.

\subsection{Modelling of hot white dwarfs}\label{ssec:model_spectra}
\begin{figure}
\includegraphics[height=\columnwidth, angle=270]{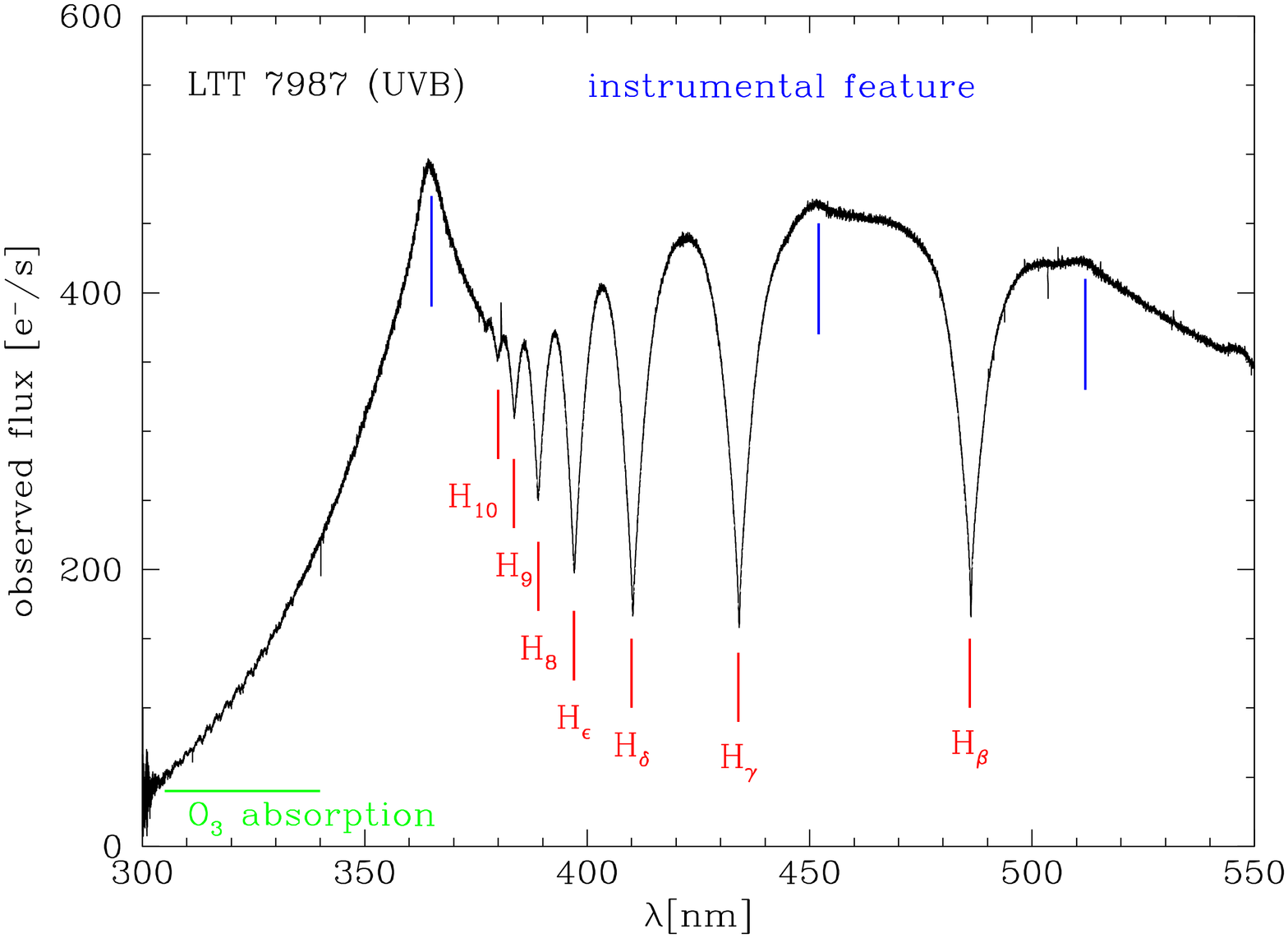}
\includegraphics[height=\columnwidth, angle=270]{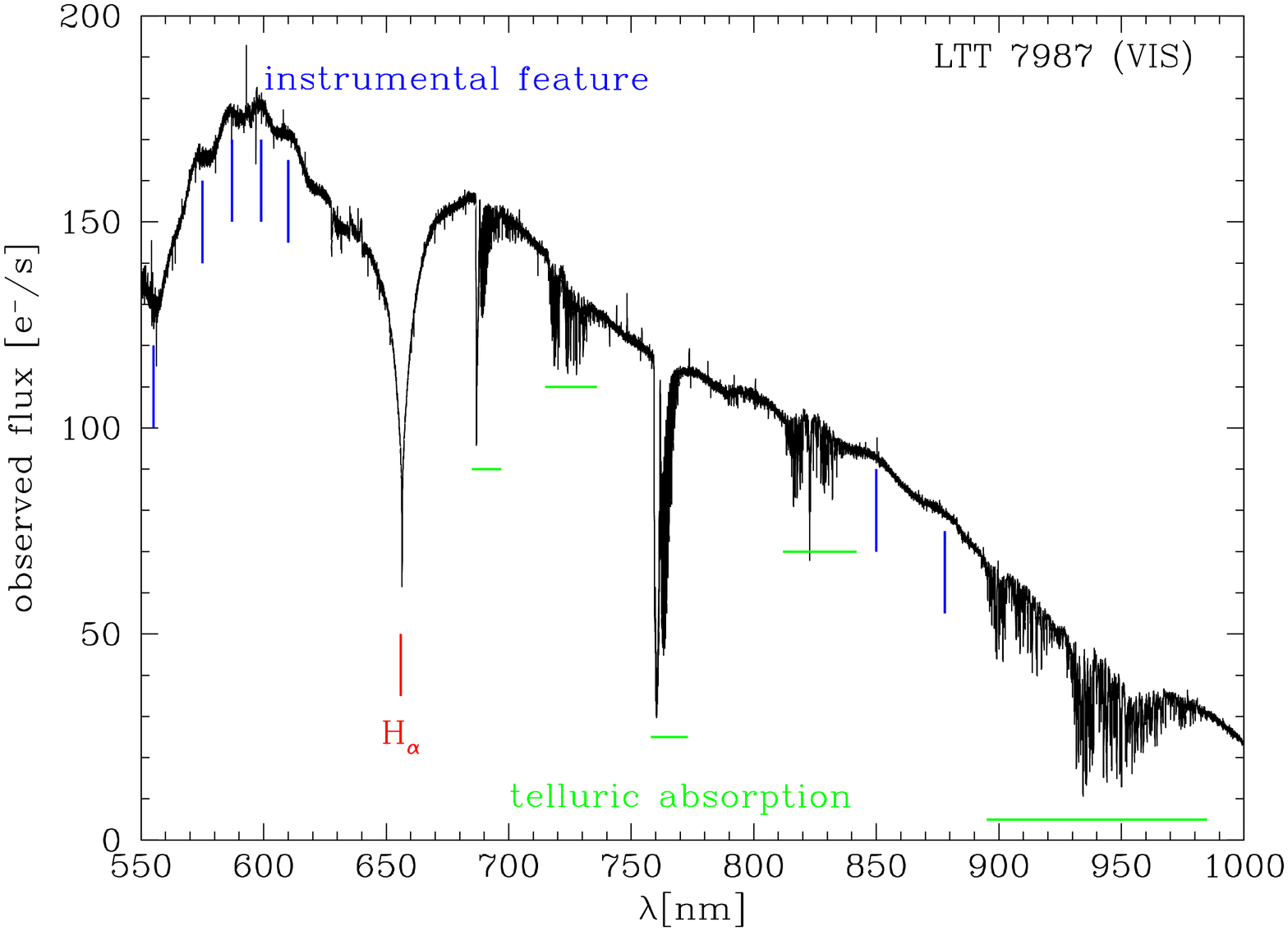}
\includegraphics[height=\columnwidth, angle=270]{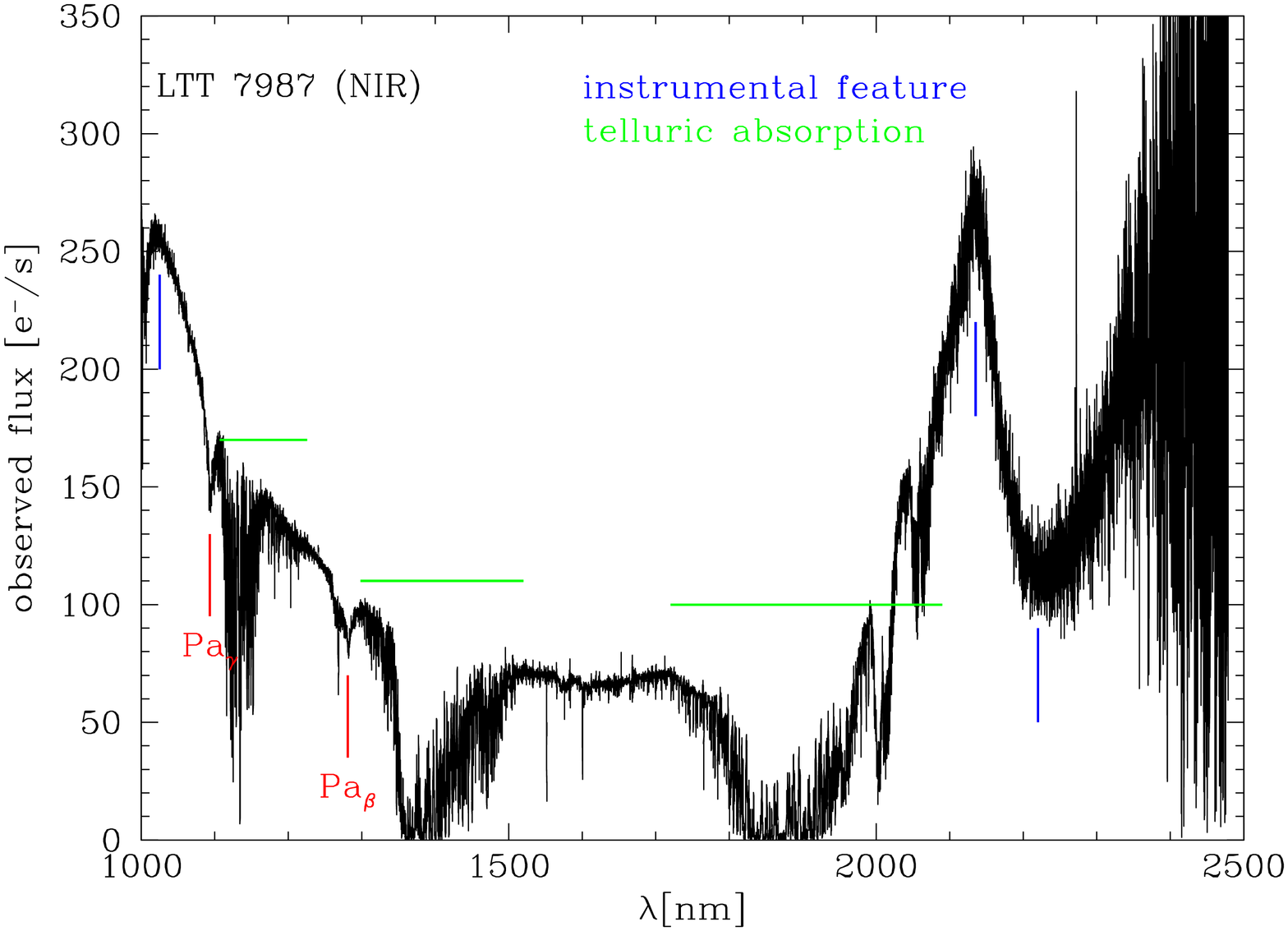}
\caption[]{Observed X-shooter spectra of LTT\,7987, corrected for
  gain, exposure time, and extinction. The stellar lines (marked in
  red) are most pronounced in the UVB arm, while the NIR arm and the
  red part of the VIS arm are strongly affected by telluric absorption
  (marked in green). All wavelength ranges also show some evidence
  of instrumental features (marked in blue).}\label{fig:LTT7987_obs}
\end{figure}

The spectra of our standard stars include hydrogen lines
(in the case of Feige\,110 also helium lines) of varying strength (see
Fig.\,\ref{fig:LTT7987_obs} for an example) that need to be properly
modelled if one wants to sample the response of an instrument on
scales of some nanometres.

For the past 20 years the physical parameters of hot white dwarfs and
other hot, high-gravity stars have been determined by fitting the
profiles of the hydrogen (and/or helium) absorption lines in their
optical spectra (\citealt{besa92}; see \citealt{gian11} and \citealt{giam12}
for more recent examples and \citealt{kona01} for work with
high-resolution \'echelle spectra).

\begin{figure*}[!ht]
\includegraphics{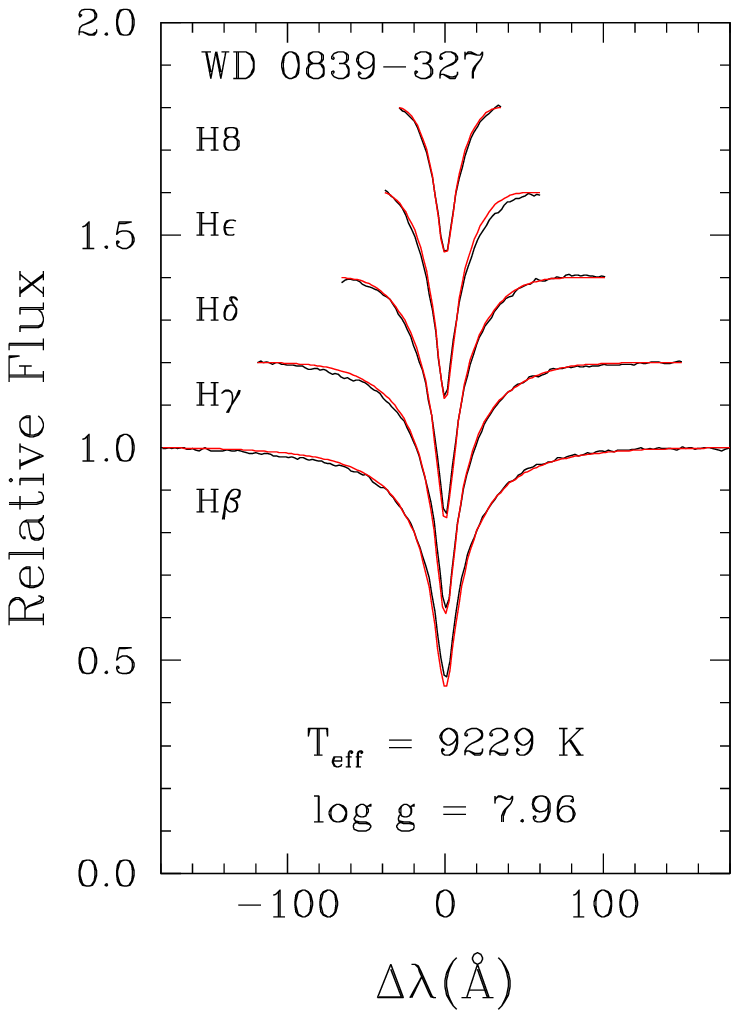}
\hspace*{6cm}
\includegraphics[height=0.6\textwidth, angle=270]{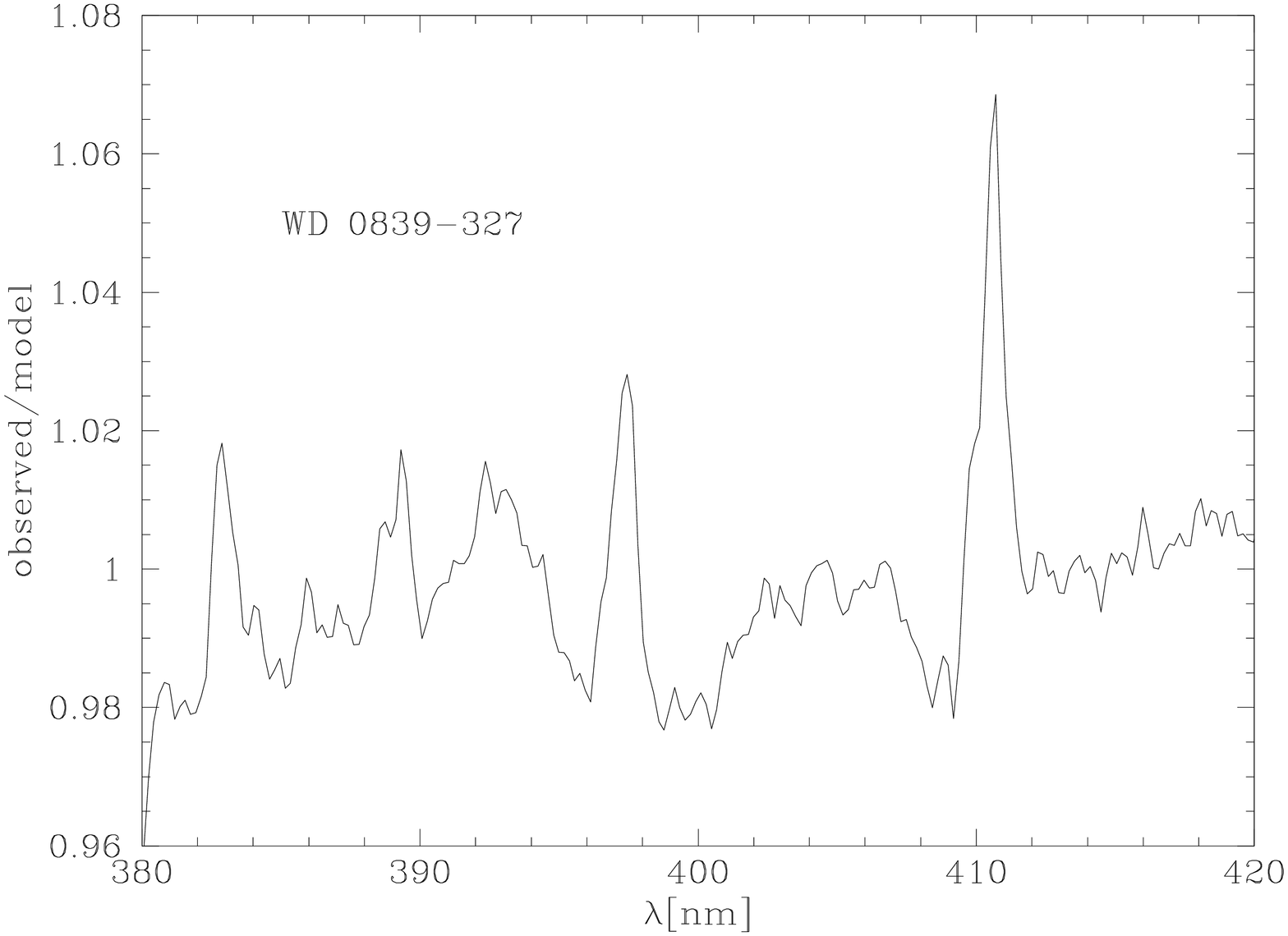}
\caption[]{Model fit for LTT\,3218 (individual lines only) from
  \citet[{\bf left}]{giam12} and the ratio between
  their observation and model spectrum ({\bf right}). The ratio plot
  on the right clearly shows residuals in the line cores and also some
  bumps between the lines.}\label{fig:Montreal}
\end{figure*}

The difficulty of modeling these spectra is illustrated in
Fig.\,\ref{fig:Montreal}.  The left panel shows an example of a parametric  fit
by \citet{giam12}. The method of fitting the
  Balmer lines profiles has the advantage that it
is sensitive to changes in temperature and surface gravity at high effective
temperature, when optical and near-infrared photometry become insensitive to
changes in these parameters. Consequently, this means that model
spectra with a wrong effective temperature or surface gravity will not
correctly describe the strong hydrogen and/or helium absorption lines in the
spectra of these stars. Such a mismatch is most severe at the blue end of the
wavelength range studied in this paper, namely between 380\,nm and 420\,nm,
where the lines may overlap with each other, causing a lack of continuum. An
example of such a mismatch can be seen in the right plot in
Fig.\,\ref{fig:Montreal}, which shows the ratio of the observed spectrum to the
fitted model spectrum for LTT\,3218. The bump between 400\,nm and 410\,nm is
caused by an imperfect description of the line overlap region between the
Balmer lines H$\delta$ and H$\epsilon$. Compared to the line depths of about
50\% the effects are small (at about 2\%), but sufficient to introduce artefacts
in the resulting response curves. To ensure that no such mismatches exist in
the finally selected reference spectra, however, one has to know the true
response of the instrument with which the spectra, that are used for the
analysis, are observed.

\begin{figure}
\includegraphics[height=\columnwidth, angle=270]{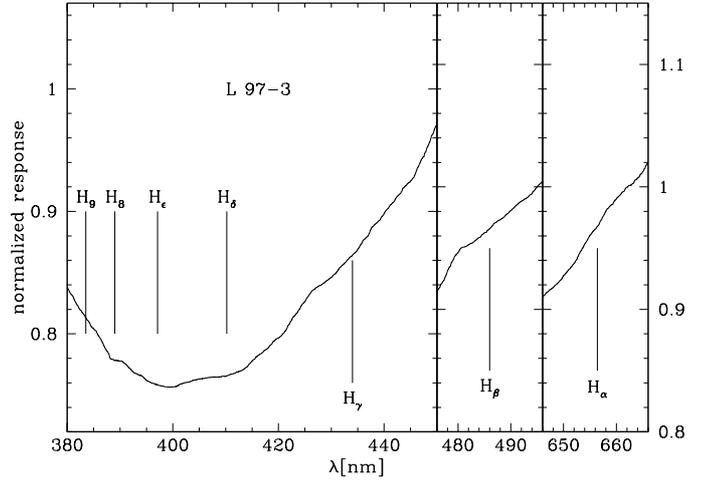}
\caption[]{Average ratios of observed standard spectrum and shifted
  reference spectrum for UVB and VIS data of L97-3 for the same
  wavelength ranges as in Fig.\,\ref{fig:new_resp}. The two narrow
  windows on the right share the same range along the y-axis. None of
  the residuals seen in Fig.\,\ref{fig:new_resp} is visible here.}\label{fig:L97_resp}
\end{figure}

By contrast, the featureless spectra of L97$-$3 can easily be modelled
at the same wavelengths, which was our primary motivation for
including it in our sample.  This is illustrated in
Fig.\,\ref{fig:L97_resp}, which shows the ratio of model spectrum to
observed spectrum from the observed spectra of L97-3 for the regions
containing strong lines in the spectra of the flux standard stars
discussed here.  We used as reference data a stellar model spectrum
calculated by \citet{koes10} for the parameters reported by
\citet{giam12} for this star. This model spectrum
was adjusted to the photometry of the star reported in
\href{http://simbad.u-strasbg.fr/simbad/sim-id?Ident=L97-3&NbIdent=1&Radius=2&Radius.unit=arcmin&submit=submit+id}{Simbad}\footnote{original data from
  \citet[$uvby$]{hame98}; \citet[$RI$]{suhe07}; \citet[2MASS point source catalogue, $JHK$]{skcu06}.} using the
pseudo-magnitudes described in Sect.\,\ref{ssec:phot_cal}.  The curves
in Fig.\,\ref{fig:L97_resp} are rather smooth without significant
structure on nanometre scales  (as opposed to
  Fig.\,\ref{fig:Montreal}). This suggests that the
spectrum has been sucessfully modelled, and the structure in the ratio
of observed and modelled spectra reflects the true response curve of the
instruments.

\begin{table}
\caption[]{Parameters for the flux standard star reference spectra}\label{tab:par}
\begin{tabular}{lrlcc}
\hline
\hline
Star & \multicolumn{4}{c}{Selected Reference Spectrum}\\
& $T_{\rm eff}$ & $\log g$ & Reference & Model \\
& [K]  & [cm s$^{-2}$] & & \\
\hline
EG\,274    & 25985 & 7.96 & 2 & 2 \\
Feige\,110 & 45250 & 5.8 & 3 & 4 \\
LTT\,3218  & 9081 & 7.71 & 2 & 2 \\
LTT\,7987  & 16147 & 7.98 & 2 & 1\\
GD\,71    & 33590 & 7.93 & 5 & 5\\  
GD\,153   & 40320 & 7.93 & 5 & 1\\
\hline 
L97$-$3 & 10917 & 8.15 & 2 & 1 \\ 
\hline 
\multicolumn{5}{l}{(1) \citet{koes10}; (2) \citet{giam12};}\\
\multicolumn{5}{l}{(3) this paper; (4) TMAP; (5) \citet{gian11}.}
\end{tabular}
\end{table}

\subsection{Selection of model spectra}\label{ssec:method}

We used the X-shooter observations described in Sect.\,\ref{ssec:data} to
verify the quality of the various model spectra available for our
  standard stars by comparing instrumental response curves from
  different standard stars. We concentrated
first on the wavelength range 380\,nm to 420\,nm for the reasons
discussed above. Once an acceptable description of the blue range has
been achieved we will verify its suitability for the redder
wavelength ranges as well. 

For our analysis we used one-dimensional extracted and merged
X-shooter spectra of the flux standard stars processed with the reflex
workflow\footnote{\url{http://www.eso.org/sci/software/pipelines/}}
\citep{frro13} of X-shooter. These spectra were then corrected for
atmospheric extinction using the Paranal extinction curve from
\citet[see also Sect.\,\ref{ssec:ext_300nm}]{pamo11} and normalized by
exposure time.

We shifted the model spectrum for a given star to the same radial
velocity as the observed spectrum to avoid the introduction of
pseudo-P\,Cygni profiles when the observed spectrum is divided by the
model spectrum. We resample the noise-free model spectra instead of
the observed spectra because this can be done without loss of
information. The radial velocity was obtained from the positions of
the line cores of H$\delta$ (UVB), H$\alpha$ (VIS), and Pa$\gamma$
(NIR). Each observed standard star spectrum was divided by the shifted
reference star spectrum. Finally the ratios were averaged per star to
achieve a better signal-to-noise ratio. This provided us with raw,
i.e. unsmoothed, response curves that allowed us to look for
systematic discrepancies between the observed and the model spectra. A
response curve is instrument specific and therefore expected to be
stable for a long time. Any feature in these ratios that appears only
for a particular standard star points strongly towards deficiencies in
the reference spectra of these standard stars. Instrumental effects
should not depend on which standard star is used. We selected as best
description of the UVB spectrum those model spectra that showed the
smallest star-specific bumps in the averaged ratio
spectra. Section\,\ref{ssec:bump_UVB} describes how we adjusted the
selected model spectra empirically to remove the remaining small
discrepancies.

\subsubsection{White dwarfs}

For the purpose of this work, we considered the models of \object{EG
  274}, \object{LTT 3218}, and \object{LTT 7987} presented by
\citet{giam12} and the models of \object{GD 71} and \object{GD 153}
  by \citet{gian11}, hereafter referred to as Montr\'eal spectra,
and the model spectra calculated by \citet[in the following referred to
  as Kiel spectra]{koes10} for the parameters from \citet{giam12}
and \citet{gian11}.

These model spectra provided good fits to all white dwarfs, but bumps
like the one seen in Fig.\,\ref{fig:Montreal} can be seen in the
ratios of observed to stellar model spectra, pointing to remaining
mismatches.  Residuals of this kind (of the order of 2\%-3\%) are not
unexpected for stars with such strong (flux in line core
about 40\% of continuum level) and overlapping lines (line widths up
to 20\,nm as can be seen in Fig.\,\ref{fig:Montreal}, left
plot). 

For EG\,274, GD\,71, and LTT\,3218 the Montr\'eal model spectra showed
the smallest star-specific bumps in the averaged ratio spectra, while
the Kiel model spectra for the Montr\'eal parameters provided the best
description for GD\,153 and LTT\,7987.  The parameters are listed in
Table\,\ref{tab:par}.
\subsubsection{Feige\,110}

For \object{Feige\,110}, we fit FORS2\footnote{ESO's FOcal
  Reducer/low dispersion Spectrograph 2, which covers the wavelength
  range 320\,nm -- 1\,100\,nm at various resolutions} and X-shooter
UVB spectra (flux-calibrated with a response curve obtained from
  GD\,71 observations) with TMAP (\href{http://astro.uni-tuebingen.de/~TMAP}{T\"ubingen NLTE Model-Atmosphere
Package}) model spectra \citep{rade03,wede03}
to obtain an appropriate reference spectrum.

\subsection{Correction of model spectra at short wavelengths}\label{ssec:bump_UVB}
\begin{figure}
\includegraphics[width=\columnwidth]{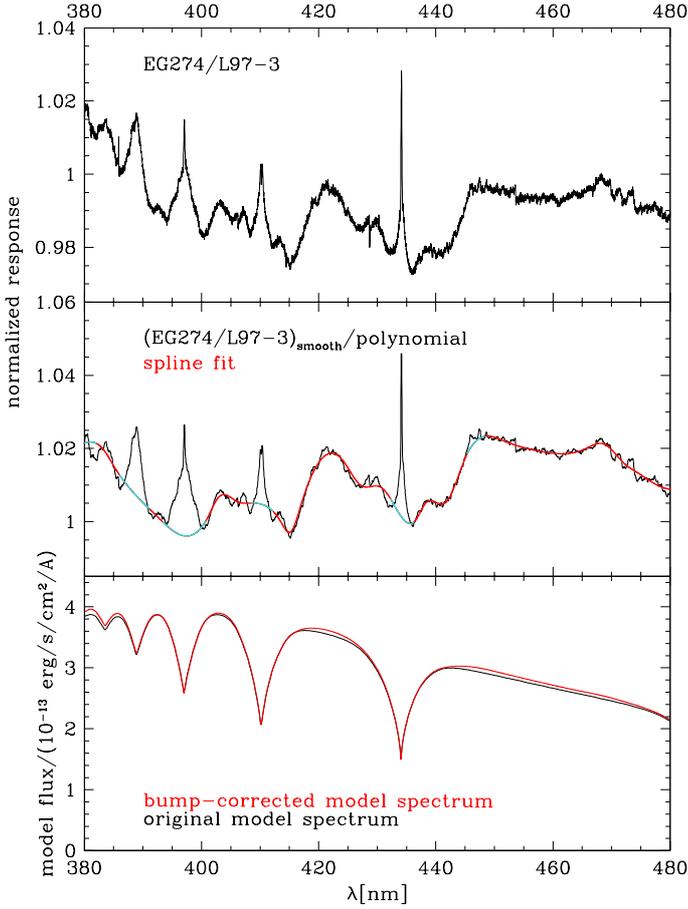}
\caption[]{{\bf Top:} Average raw response ((extinction-corrected
  observation)/(radial-velocity shifted model)), divided by the
  smoothed response for L97-3, for EG\,274. The narrow spikes are due
  to mismatches in the line cores, while the broad bumps at 405\,nm
  and 420\,nm are due to imperfect line broadening, which
  underestimates the flux between the lines in the model spectra. {\bf
    Middle:} Raw response from the top plot, with large scale
  variations removed by a low-order polynomial fit and smoothed
  (black). The black curve was fit with a spline (red) after masking
  the regions of the line cores (masked regions in cyan). {\bf
    Bottom:} Model spectrum before correction (black) and corrected
  with the fit (red). }\label{fig:EG274resm_L97n}
\end{figure}

In order to correct the mismatches between model and observed spectra
that cause the bumps we reprocessed all flux standard stars observed
in NODDING in 2012 with the new model spectra as reference spectra. We
first divided each (extinction-corrected) one-dimensional spectrum by
the radial-velocity shifted reference spectrum and then by the
smoothed response curve derived from L97-3. The products of this
procedure are expected to be equal to 1 in an ideal case, but
differences in atmospheric conditions result in residual slopes. The
parking of the Atmospheric Dispersion Compensators\footnote{The ADCs
  in the UVB and VIS arm minimise slit losses, keep the target at the
  same position, and allow observations at any position angle of the
  slit on the sky up to zenith distance of 60$^{\circ}$} since August
2012 might also play a role here, as the UVB spectra observed at
higher airmass are no longer at a constant position along the slit at
all wavelengths, but show instead some curvature at shorter
wavelengths, which may in turn affect the extraction. In addition,
remaining mismatches between model spectra and observations cause
narrow spikes and bumps that we want to correct (see
Fig.\,\ref{fig:EG274resm_L97n}, top).  For every star we fitted the
shape of this ratio spectrum by a low-order polynomial fit to remove
the large-scale variations. Then we smoothed the result and fit the
remaining bumpy regions (see Fig.\,\ref{fig:EG274resm_L97n}, middle,
for an example). Applying the fit to the model spectrum we obtained
the red curve shown in Fig.\,\ref{fig:EG274resm_L97n} (bottom).

The bump corrections were smallest for GD\,71, GD\,153, and Feige\,110,
while EG\,274, LTT\,7987, and LTT\,3218 had significant corrections
(increasing in this order). This is expected as the last three stars
have the strongest and widest lines and thus present the biggest
challenge when it comes to the correct treatment of overlapping lines.
Since the mismatches in the line cores were not corrected (as they
vary with spectral resolution) some masking is still required when
fitting a response (see Sect.\,\ref{sec:pipe}). We also verified that
in the spectral range 550\,nm to 2\,500\,nm the only residuals seen
are those in the line cores of hydrogen and/or helium lines.

\subsection{Absolute flux calibration}\label{ssec:phot_cal}

\begin{figure}
\includegraphics[height=\columnwidth, angle=270]{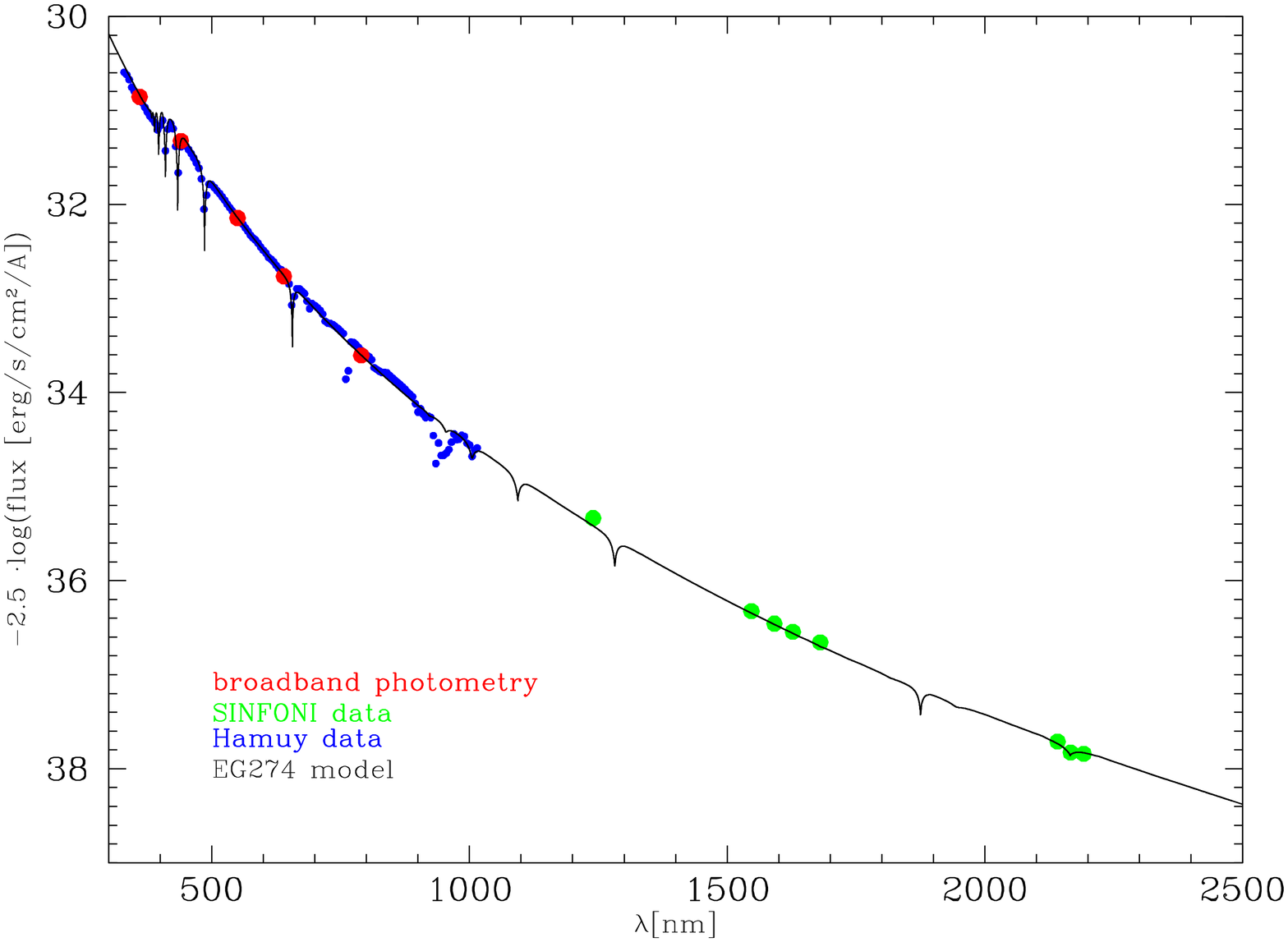}
\caption[]{Model spectrum for EG\,274 (black), aligned to the
    data from \citet[blue]{hamuy92,hamuy94},
    \citet[green]{veke08}, and broadband photometry from
    the SIMBAD database (red).}\label{fig:align}
\end{figure}

We have now defined a set of model spectra which adequately reproduce
the observed X-shooter spectra of the six flux standard stars, with residuals
(outside line cores, that are affected by resolution effects) well
below 2\%. To allow a proper flux calibration we have to verify that
the overall flux distribution of these model spectra reproduces the
independently observed (spectro-)photometric data. To do so we
convolved the model spectra to a resolution of 1.6\,nm for the UVB/VIS
spectral range and then binned them to 5\,nm steps to reproduce the
\citet{hamuy92,hamuy94} \href{http://vizier.u-strasbg.fr/viz-bin/VizieR?-source=II/179}{data}. These reference data are based on absolutely flux
  calibrated low-resolution spectra, that are tied to the Vega
  calibration of \citet{haye85} via a recalibration of the
  secondary standard stars from \citet{tayl84}. For the NIR
range we convolved the data to a resolving power of 2000 ($J$ band)
and 1500 ($H+K$ band) and integrated them over the wavelength ranges
given in \citet{veke08} for EG\,274, Feige\,110,
LTT\,3218, and LTT\,7987. These flux values were then converted to
pseudo-magnitudes
\begin{equation}
mag = -2.5\cdot\log (flux)
\end{equation}
and aligned to the respective Hamuy and
SINFONI (Spectrograph for Integral Field Observations in the Near
  Infra-red at the ESO VLT; \citealt{eiab03}) data by a
constant factor across the full wavelength
range. Figure\,\ref{fig:align} shows the results for EG\,274. 

\begin{figure}
\includegraphics[height=\columnwidth, angle=270]{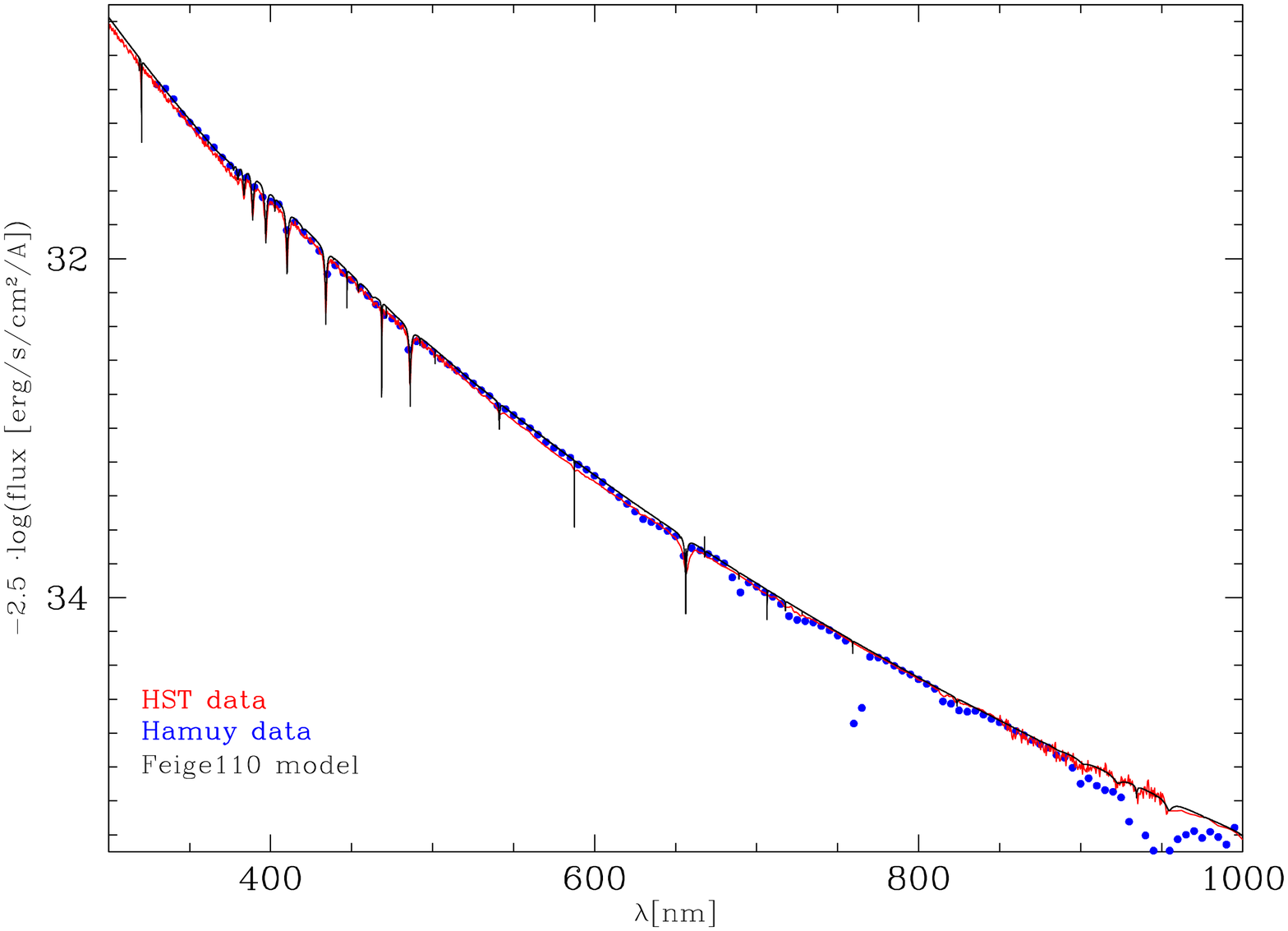}
\caption[]{Aligned model spectrum for Feige\,110 (black) compared to the
    data from \citet[blue]{hamuy92,hamuy94} and
    from HST (red).}\label{fig:F110_align}
\end{figure}

For GD\,71 and GD\,153 our new model spectra could also be aligned by
a constant factor to the HST model spectra, which are tied to the
  Vega scale by HST STIS and NICMOS observations \citep{bohlin01}. In
  Fig.\,\ref{fig:F110_align} we compare our aligned model spectrum for
  Feige\,110 to the data from Hamuy and from HST.

\begin{figure*}
\includegraphics[height=\textwidth, angle=270]{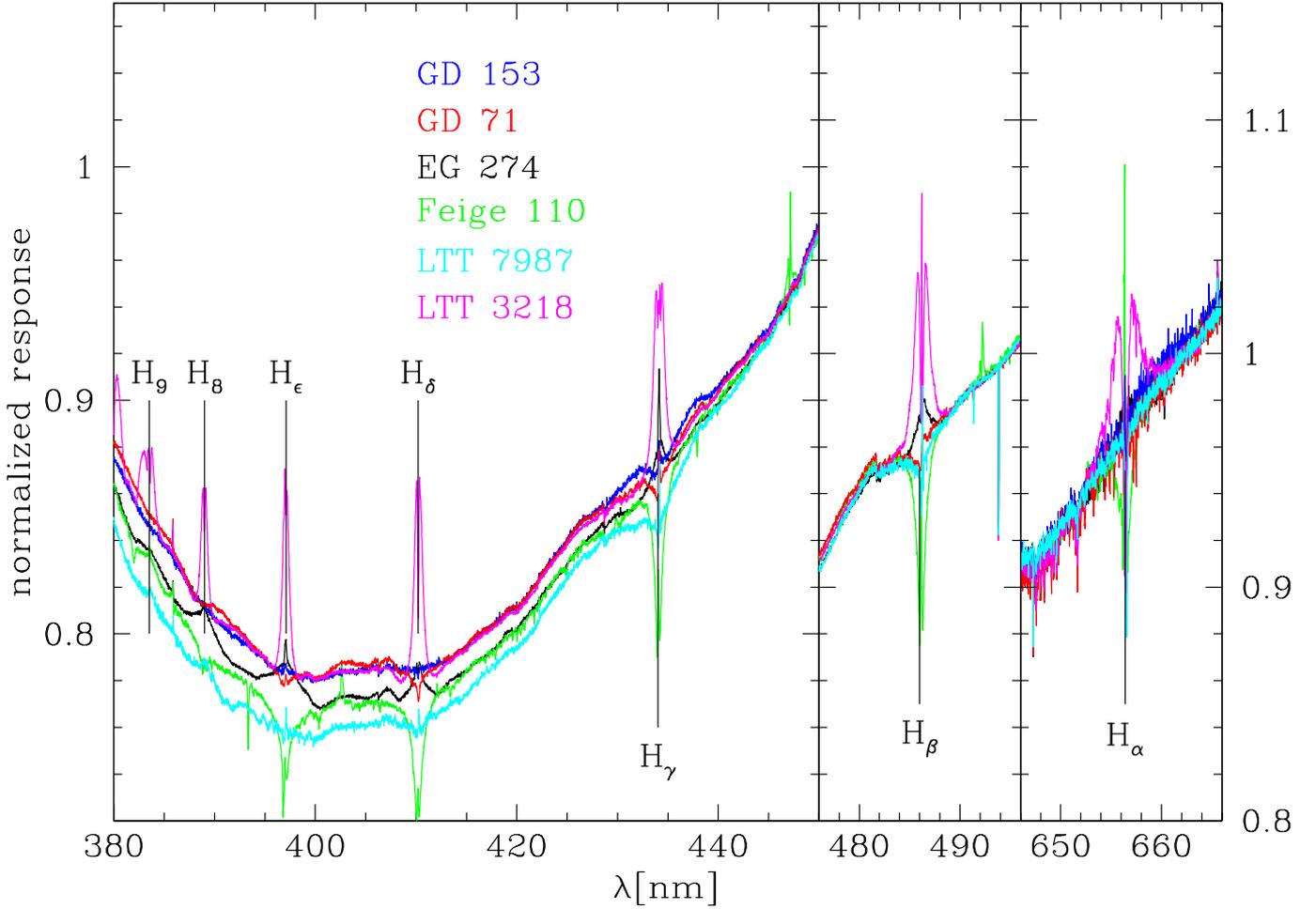}
\caption[]{Average ratios of observed standard star spectrum 
    (corrected for atmospheric extinction) and
  shifted reference spectrum for UVB and VIS data for the regions
  containing Balmer lines, using the bump-corrected
  model spectra. The curves have been normalized at the red end of the
  respective plot windows, which show the Balmer lines H$\alpha$
  (right), H$\beta$ (middle), and H$\gamma$ to H$9$ (left). The
    two narrow windows on the right share the same range along the y-axis.
}\label{fig:new_resp}
\end{figure*}

Figure\,\ref{fig:new_resp} shows the
average ratio of observed and reference spectra for the new model
spectra.

\section{Correction of atmospheric effects}\label{sec:atmos}

We have now defined reference spectra that describe the flux standard
stars very well. In order to derive a response curve the effects of
the Earth's atmosphere have to be removed from the observed
spectra. Otherwise the response curve will contain a mixture of
instrumental effects and atmospheric effects.  The atmospheric effects
consist of two major parts: atmospheric extinction (small
variation with time in the case of Paranal due to the low
aerosol content, which governs the extinction variability) and
telluric absorption lines (strong variation with time in the case of
water vapour). The first affects principally the wavelength
range 300\,nm to 1\,000\,nm, while the second is important for data
above 680\,nm (X-shooter NIR arm and the redder part of the VIS arm;
see Fig.\,\ref{fig:LTT7987_obs}, middle and bottom plot).

\subsection{Correcting the atmospheric extinction down to
  300\,nm}\label{ssec:ext_300nm} 

\begin{figure}
\includegraphics[height=\columnwidth, angle=270]{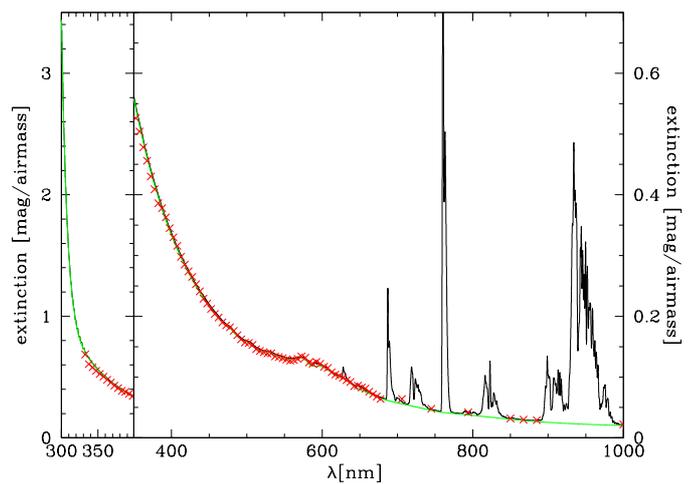}
\caption[]{LBLRTM model for Paranal (black line) compared to
    the FORS2 data points from \citet[red
    crosses]{pamo11} and the interpolated curve (green line). 
}\label{fig:atm_ext}
\end{figure}

\citet{pamo11} provide an extinction curve for Paranal that is based
on FORS2 observations and therefore covers only wavelengths longer
than 320\,nm. X-shooter data, however, extend down to 300\,nm. In
order to cover a larger wavelength range we obtained from F. Patat a
Line-By-Line Radiative Transfer Model
(\href{http://rtweb.aer.com/lblrtm.html}{LBLRTM}, \citealt{clsh05})
spectrum for Paranal, that describes the continuous and line
absorption caused by the Earth's atmosphere. As the spectrum was
calculated without aerosol contributions, we added the aereosol
contribution defined as $k_{\rm aer} = 0.014 \cdot (\lambda [\mu
  m])^{-1.38}$ (see \citet{pamo11} for more details). The
LBLRTM model spectrum was calculated for the wavelength range
300\,nm--1099.4\,nm and also contains absorption by telluric lines,
which vary rapidly with time and can therefore not be corrected by a
static extinction curve (see Sect.\,\ref{ssec:telluric} for the
correction of telluric lines). To exclude the telluric lines from the
extinction curve we interpolated across the regions of strong telluric
absorption in the VIS range (585.5\,nm--599.2\,nm,
626.1\,nm--634.9\,nm, 643.8\,nm--660.0\,nm, 682.1\,nm--709.4\,nm,
712.7\,nm--743.4\,nm, 756.2\,nm--773.1\,nm, 780.1\,nm--861.3\,nm,
879.8\,nm--1033.8\,nm, $>$1050\,nm). Figure\,\ref{fig:atm_ext} shows the
model and our interpolation in comparison to the FORS2 measurements.

\subsection{Telluric absorption}\label{ssec:telluric}

If one does not correct at least those parts of the spectrum that
contain low to medium telluric absorption, one has to interpolate the
response across very wide wavelength ranges, which results in
large systematic uncertainties for the resulting flux
calibration. Traditionally such corrections use so-called telluric
standard stars, i.e. stars with either no features or extremely
  well-known features in the regions of telluric absorption, that
allow the user to determine the telluric absorption spectrum (see for
instance \citealt{vacu03,mari96}). This method relies on the
assumption that the conditions governing the telluric absorption have
not changed between the observation of the science target and the
observation of the standard star. Since this assumption is often not
fulfilled, we decided instead to use a catalogue of telluric model
spectra \citep{noka12,jone13} to correct the telluric absorption
lines.  Alternatively, a large number of telluric standard star
observations can be used to extract the principal components of the
telluric transmission \citep{chen14}.

The high-resolution transmission spectra ($R$ = 60\,000) were
calculated for the wavelength range from 300\,nm to 30\,$\mu$m
using the LBLRTM radiative transfer code, the HITRAN molecular line
database (see \citealt{ROT09}), and different average atmospheric
profiles (pressure, temperature, and molecular abundances) for Cerro
Paranal. The atmospheric profiles were derived from the MIPAS
equatorial standard profile (J. Remedios 2001; see \citealt{SEI10}),
profiles for pressure, temperature, and relative humidity from the
\href{http://ready.arl.noaa.gov/gdas1.php}{Global Data Assimilation
  System (GDAS)}, provided on a 3\,h basis for a $1\degr \times
1\degr$ global grid up to an altitude of 26\,km, and data from the
Cerro Paranal meteorological station. The spectra we used cover the
airmass values 1.0, 1.5, and 2.0 at six different bimonthly periods
(plus an annual mean) and three night-time periods (plus a nocturnal
mean). We will refer to this telluric library as the time-dependent
library.

To identify the best telluric model spectrum to correct a given
observation, the telluric model spectra first have to be aligned
to the observed spectrum in both resolution and wavelength. Then
  the telluric model spectrum that leaves the smallest residuals
after applying it to the observation has to be identified. 
  We used the following step-by-step procedure:
\begin{enumerate}
\item Convert the wavelength scale (in nm) of the model and
  observed spectra to natural logarithm.
\item Extract appropriate ranges for cross correlation (NIR: 7.0$\le \ln
  \lambda_{\rm nm} \le$7.1, H$_2$O feature; VIS: 6.828$\le \ln
  \lambda_{\rm nm} \le$6.894, H$_2$O feature).
\item Cross correlate the extracted telluric model and observed spectra with a
  maximum shift of 500 sampling units (each of size $10^{-5}$)
    in natural logarithmic space.
\item Fit a Gaussian profile within $\pm$0.001 in $\ln(\lambda)$
  around the cross correlation peak.
\item Shift the logarithmic model wavelength scale by the fitted peak
  position of the cross-correlation curve.
\item Create a Gaussian with the measured FWHM.
\item Convolve the model spectrum with the Gaussian in $\ln(\lambda)$
  space (X-shooter has a constant resolving power $R$ of
    4350/7450/5300 for an 1\farcs0/0\farcs9/0\farcs9 slit in the
    UVB/VIS/NIR arm, respectively)\footnote{Since the resolution of
      the observed spectra is much lower than the resolution of the
      telluric model spectra, we assume here the resolution of the model
      spectra to be infinite.}.
\item Convert the shifted and convolved model spectrum to linear
  wavelength space.
\item Divide the observed spectrum by the convolved and shifted
  telluric model spectrum (whereby we avoid resampling the observed spectrum).
\item Use predefined continuum points (avoiding regions of strong
  telluric absorption as well as known lines of the observed star, see
  Sect.\,\ref{ssec:model_spectra}) and fit a cubic spline to the
  telluric corrected spectrum.
\item Divide the telluric corrected spectrum by the fit of the continuum.
\item Determine mean and rms for regions of moderate telluric absorption
  in the normalized corrected spectrum.
\end{enumerate}

\begin{figure}
\includegraphics[width=\columnwidth, angle=0]{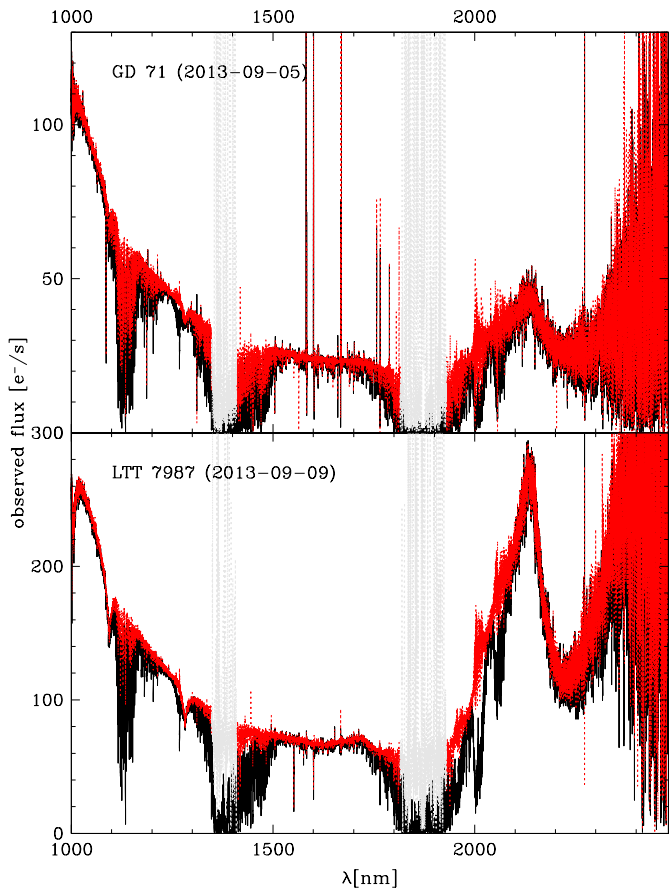}
\caption[]{Results of the telluric correction (red) for a bad
    (top) and a good (bottom) case. The grey areas mark regions of
    extremely high telluric absorption and thus high noise in the
    corrected data. The lines at 1093.5\,nm and 1281.4\,nm are stellar
    lines and the variable peak at about 2100\,nm is caused by
      variations in the flux level and spectral energy distribution of
     the NIR flat field lamp.}\label{fig:NIR_tell}
\end{figure}

Then, we choose as best fitting model the one for which the average
residuals computed over regions of moderate telluric absorption is
minimal. The telluric corrected spectrum of the flux standard star is
then compared to the stellar model spectrum
(see Sect.\,\ref{ssec:model_spectra} for details) to determine the
response. Figure\,\ref{fig:NIR_tell} shows examples for a good
telluric correction (bottom) and for undercorrection (top).

The absorption in the wavelength regions 1985\,nm--2030\,nm and
2037\,nm--2089\,nm is caused mainly by CO$_2$ features, whose
abundance is not varied in the telluric model spectra for a given
airmass. These regions were therefore ignored when looking for the best
fit. We do not restrict the search for the best telluric model
  spectrum in airmass, as we found that in some cases a telluric model
  spectrum for an airmass quite different from the observed one can
  still provide a good fit in the regions that we analyse.

The telluric model spectra discussed so far do not cover high
precipitable water vapour ({\bf PWV}) conditions, which result in
  stronger telluric absorption lines, and therefore spectra taken
during such conditions are undercorrected (see
Fig.\,\ref{fig:NIR_tell}, top, for an example). This problem is more
pronounced in the NIR than in the VIS arm. Therefore we
  investigated another set of model spectra, which extends to higher
  PWV conditions (PWV = 0\ldots20\,mm) and higher airmass values
  (PWV-dependent library). Tests with these models showed good
  corrections for NIR data taken under high PWV conditions (which could not
  be corrected with the time-dependent grid), but were less successful
  for data taken during intermediate and low PWV
  conditions. For the VIS arm the PWV-dependent telluric model catalogue
  provided good corrections for all of the about 100 flux
    standard spectra that we tested.

Users who want to employ the telluric model spectra to correct their
data can find information about the Cerro Paranal sky model (see
\citealt{noka12}) at
\url{http://www.eso.org/sci/software/pipelines/skytools/} together
with the following
\href{ftp://ftp.eso.org/pub/dfs/pipelines/skytools/telluric_libs}{pre-calculated
  libraries}:
\begin{enumerate}
\item PWV-dependent library with 45 spectra at airmasses
  1.0, 1.5, 2.0, 2.5, 3.0, and PWV (in mm) of 0.5, 1.0, 1.5, 2.5, 3.5,
  5.0, 7.5, 10.0, and 20.0. 
\item Latest version of time-dependent library with 35 spectra at airmasses
  1.0, 1.5, 2.0, 2.5, 3.0, and bi-monthly average PWV values 
  (1=December/January \ldots 6=October/November) as well as a yearly
  average PWV content (labelled 0). Compared to the time-dependent
  library used in this paper the airmass coverage has been increased,
  the night-time periods have been dropped, and the CO$_2$ content has
  been updated.
\end{enumerate}
The spectra are provided with resolutions of 60\,000 and 300\,000 and
do not include Rayleigh scattering.  The
\href{http://www.eso.org/observing/etc/skycalc/skycalc.htm}{SkyCalc
  web application} can always be used to produce sky radiance and
transmission spectra with more specific parameters, including Rayleigh
scattering. 

The quality of the telluric correction has to be kept in mind when
fitting a response to the ratio of such a corrected spectrum to the
corresponding stellar model spectrum, as regions with
  remaining absorption or overcorrected features may distort the
  fit. For this reason we finally decided to mask the following
regions when fitting a response curve to avoid potential telluric
residuals: 634\,nm--642\,nm, 684\,nm--696\,nm, 714.7\,nm--732.3\,nm,
757.5\,nm--770.5\,nm, 813\,nm--836.5\,nm, 893.9\,nm--924\,nm,
928\,nm--983\,nm, 1081\,nm--1171\,nm, 1267\,nm--1271\,nm,
1300\,nm--1503\,nm, 1735\,nm--1981\,nm, 1995\,nm--2035\,nm,
2048\,nm--2082\,nm.
We note that this masking rejects far less of the spectrum than would be
necessary if no telluric correction had been applied. 
\begin{figure}[!h]
\includegraphics[height=\columnwidth, angle=270]{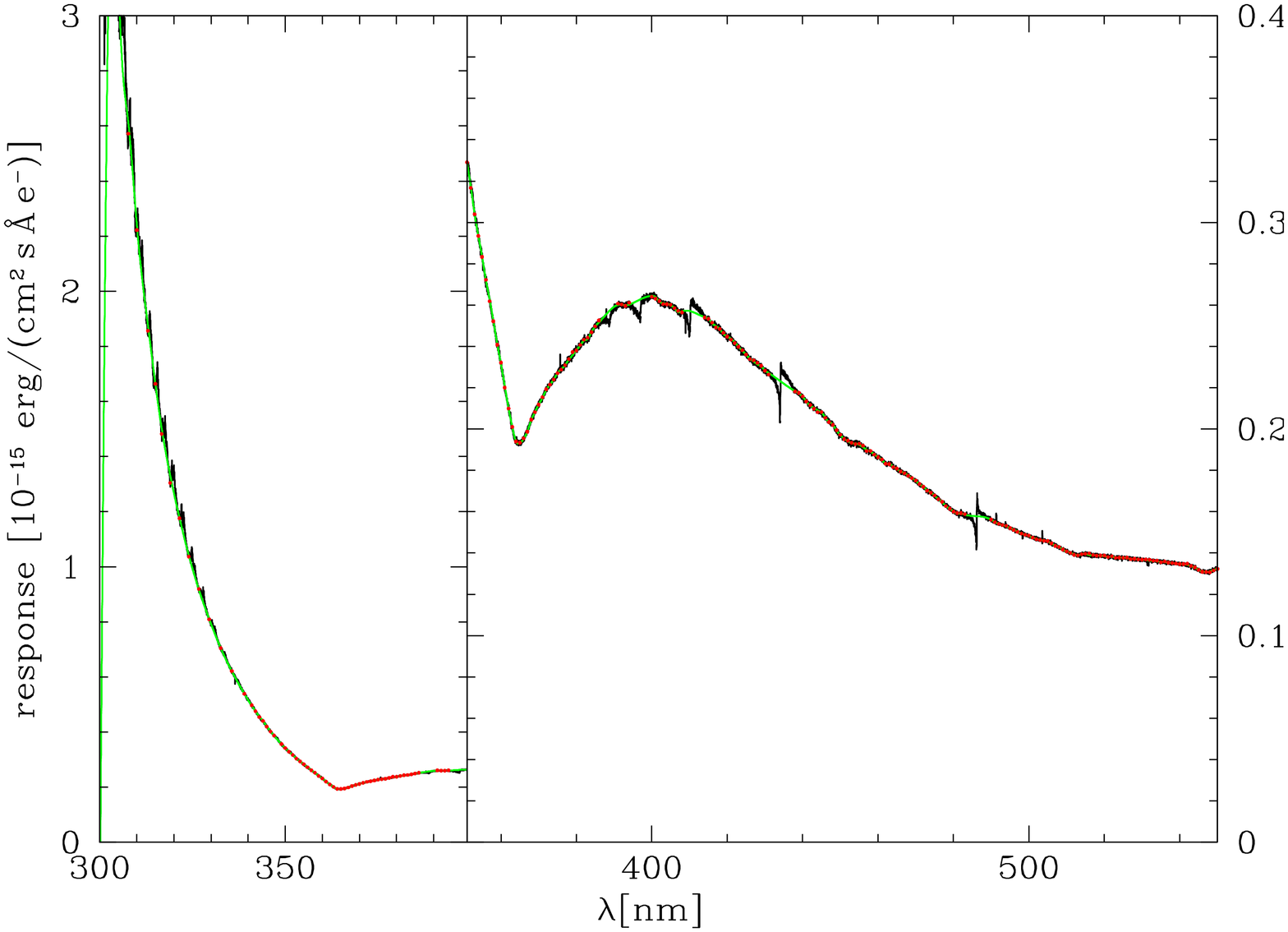}
\includegraphics[height=\columnwidth, angle=270]{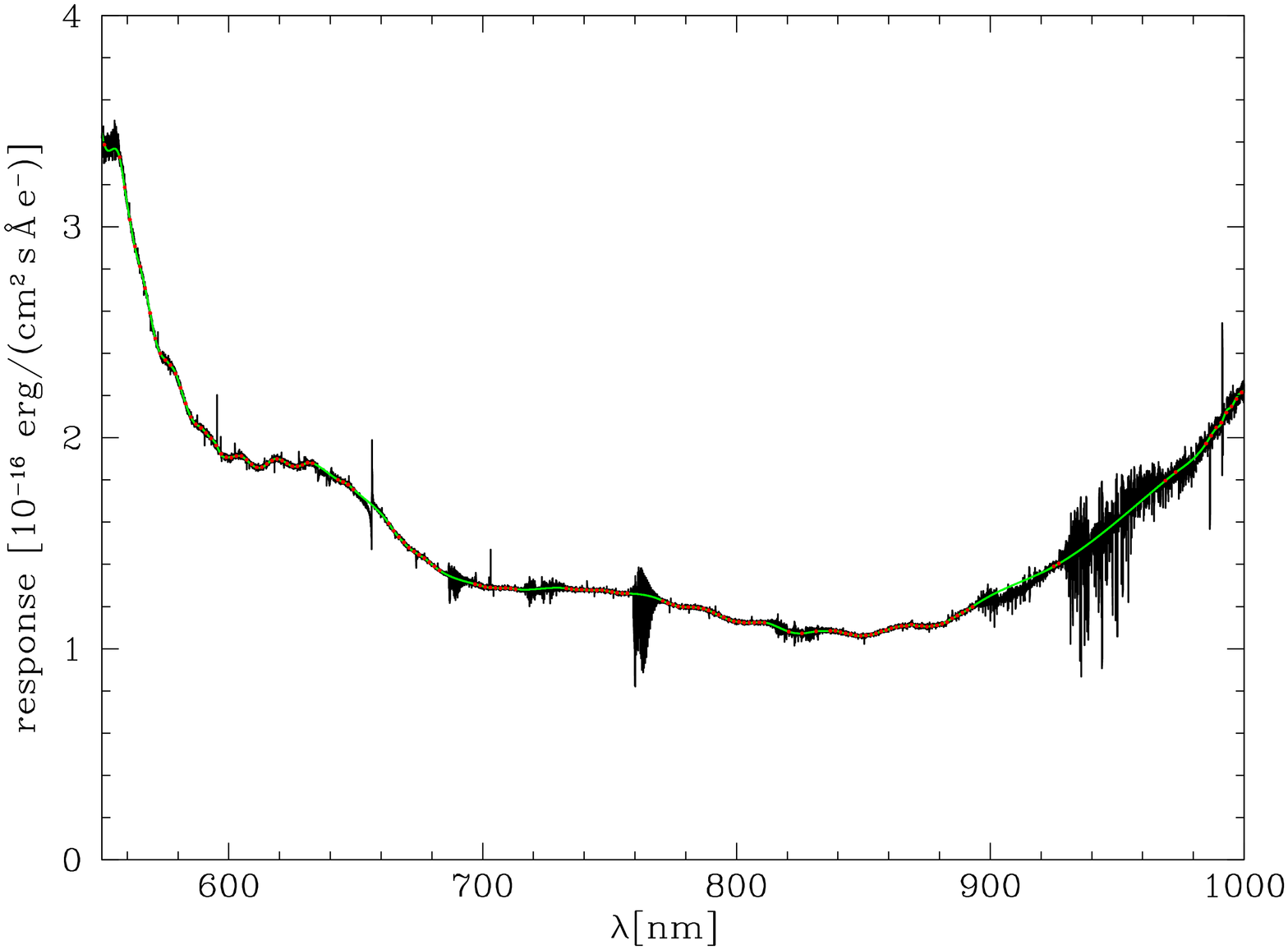}
\includegraphics[height=\columnwidth, angle=270]{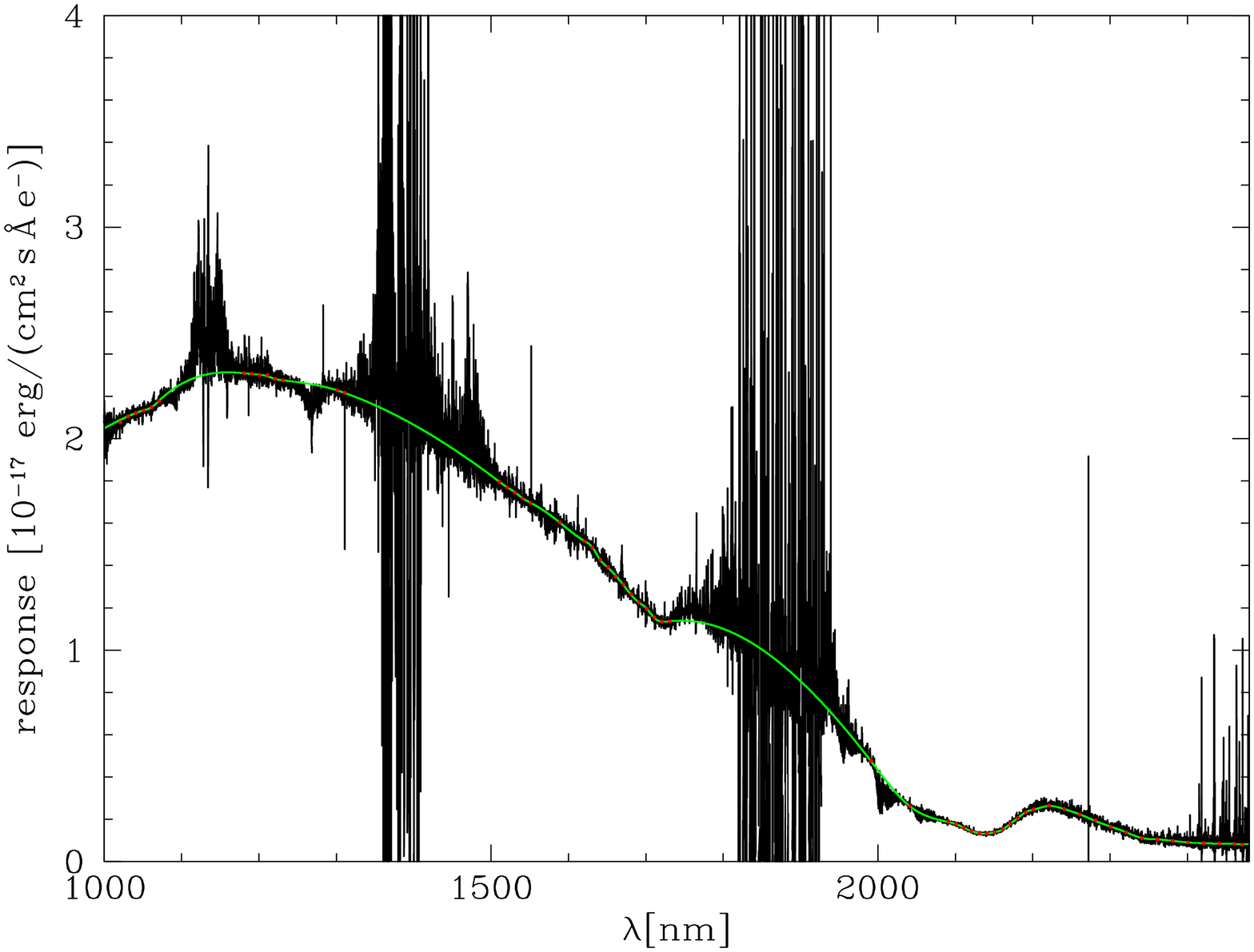}
\caption[]{X-shooter response curves derived for EG\,274. The red dots
  mark the fit points described in Sect.\,\ref{ssec:fit} and the green
  line marks the fit through these points.
}\label{fig:EG274_res}
\end{figure}

\section{Pipeline implementation}\label{sec:pipe}

The procedures described in the previous sections have been
  implemented in ESO's pipeline\footnote{http://www.eso.org/pipelines}
  for the reduction of X-shooter data.
Below we provide some more details on the algorithms and reference
data used in the pipeline.

\subsection{Radial velocity correction}
As described in Sect.\,\ref{ssec:method} the model spectra need
  to be shifted to the same radial velocity as the observed spectra to
  avoid the generation of pseudo-P\,Cygni profiles in the response curve.
The pipeline uses a robust 
polynomial fit to the line core to determine the radial velocity.  The
fit accuracy is limited to 1 pixel (0.02\,nm for UVB and VIS,
  0.06\,nm for NIR).
Because of the limited accuracy of the
radial velocity correction pseudo-P\,Cygni profiles are sometimes created when
dividing the shifted reference spectrum by the observed one. This
effect is taken into account when defining the fit regions for the response
curve (see Sect.\,\ref{ssec:fit}).

\subsection{New fitting routine for response determination}\label{ssec:fit}

The ratio of the shifted reference spectrum to the observed spectrum
(telluric corrected for VIS and NIR) shows residuals at the line cores
(see Figs.\,\ref{fig:new_resp} and \ref{fig:EG274_res}). At the same
time both UVB and VIS spectra show variations on intermediate scales
that need to be fit by the response curve because they are
  instrumental features (see
Fig.\,\ref{fig:LTT7987_obs}; features at 365\,nm and
560--620\,nm). Therefore, we created for each flux standard star and
arm a list of wavelength points at which the response is fit with a
cubic spline fit. The pipeline determines at each point the median of
the ratio over a predefined range ($\pm$0.04\,nm for UVB, $\pm$1\,nm
for VIS, and $\pm$5\,nm for NIR). The points are selected so that the
wavelength ranges used for the median do not overlap and regions of
strong telluric absorption (e.g. between $J$ and $H$ and between $H$
and $K$) are excluded (see Sect.\,\ref{ssec:telluric} for
details). Figure\,\ref{fig:EG274_res} shows the response curves derived
for EG\,274 with the pipeline. This new routine has been implemented
in the X-shooter pipeline since version 2.2.0.

\begin{figure*}[!ht]
\includegraphics[height=\textwidth, angle=270]{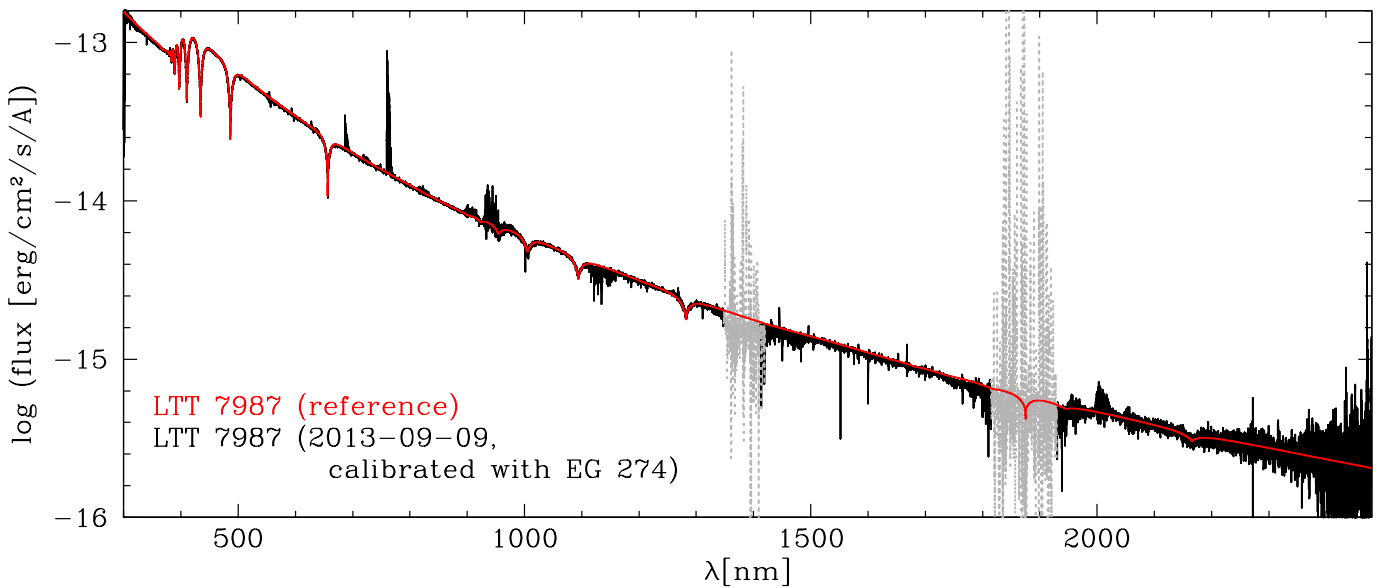}
\caption[]{Spectrum of LTT\,7987, corrected for telluric absorption
  and flux-calibrated with a response curve derived from observations
  of EG\,274. The red curve is the reference spectrum for
  LTT\,7987. Regions of very high telluric absorption are marked in
  grey.}\label{fig:cross_cal}
\end{figure*}

\subsection{New reference data}\label{ssec:pipe_ref} 
The newly derived reference spectra for the flux standard stars
described in Sect.\,\ref{ssec:model_spectra} have been delivered with
the X-shooter pipeline since version 2.2.0. 

The adjusted LBLRTM model is used to correct the atmospheric
extinction of X-shooter UVB and VIS spectra in pipeline version 2.3.0
and higher. For the NIR arm there is no extinction correction.

Since the PWV-dependent grid provides good corrections for NIR data
taken during high PWV conditions (which could not be corrected with
the time-dependent telluric library), but is less successful
for intermediate and low PWV conditions, we decided to use the
time-dependent telluric library described in Par.\,2 of 
  Sect.\,\ref{ssec:telluric} as pipeline default for the NIR
arm and provide the PWV-dependent library as possible alternative.
For the VIS arm the PWV-dependent telluric library is used. The
telluric correction and the corresponding catalogues have been part of
the X-shooter pipeline since version 2.2.0.

\section{Summary}\label{sec:summary}

In this paper we have presented flux-calibrated model spectra of
southern sky spectral photometric standards in the wavelength range
from 300\,nm to 2500\,nm. The calibrated model spectra are available
at the CDS. The consistency of the set of models were verified using
X-Shooter observations, and the region of strong line overlap between
380\,nm and 450\,nm in the model spectra was adjusted taking
advantage of X-shooter observations of the featureless spectrum of
L97--3. Our analysis shows that the use of stellar model spectra as
reference data for flux standard has significant advantages when it
comes to calibrating medium-resolution data as they are available on
a finely sampled grid. Another significant advantage is the fact that
model spectra are not affected by telluric absorption.

Our models are useful to determine the response curve of
spectrographs, and the resulting response curve can in turn be used to
flux-calibrate spectra of other targets. We have discussed in detail
a methodology to carry out such a flux calibration.  Beyond the proper
choice of stellar model spectra, an appropriate correction of telluric
absorption features is essential to flux calibrate data with
wavelengths above 680\,nm. Our method includes the use of telluric
model spectra to correct spectral regions with low to moderate
telluric absorption.

To illustrate the quality that can be achieved with the methods and
model spectra described in this paper, we show in
Fig.\,\ref{fig:cross_cal} an example of a standard star
(LTT\,7987). The spectrum was flux-calibrated with the response curve
determined from a different star observed on the same night, namely
EG\,274. The response curve used for this calibration is shown in
Fig.\,\ref{fig:EG274_res}.

While our set of standards is adequate to flux calibrate spectra
  in the whole southern hemisphere, additional standards would be
  useful both to enlarge the number of potential calibrators in the
  south, and to provide a similar set of standards in the northern
  hemisphere. Our methodology can be used to derive such a set of
  model spectra.  Necessary ingredients for such an endeavour are
  accurate model spectra\footnote{For hot (pre-) white dwarfs like the
    flux standard stars discussed here and many others, a good starting
    point for model spectra would be the TMAP spectral energy
    distributions at the registered VO service
    \href{http://dc.g-vo.org/theossa}{TheoSSA} provided by the
    \href{http://www.g-vo.org}{German Astrophysical Virtual
      Observatory}.}, availability of flux information (e.g. \citealt{hamuy94}), and observed uncalibrated spectra for both the new
  calibrators and at least one of the standard stars discussed here to
  correct for instrumental features.

\begin{acknowledgements}
This research has made use of NASA's Astrophysics Data System
Bibliographic Services and of the SIMBAD database, operated at CDS,
Strasbourg, France. It has been partly carried out in the framework of
the Austrian ESO In-kind project funded by the Austrian Federal
Ministry for Science and Research BM:wf under contracts
BMWF-10.490/0009- II/10/2009 and BMWF-10.490/0008-II/3/2011. We thank
D. Koester for making his model spectra available to us. We
appreciate J.\,Pritchard's comments that improved the readability of
this paper. We are grateful to J.\,Vernet, V.\,Mainieri, and
F.\,Kerber for sharing their SINFONI results with us.  We thank
F.\,Patat, R.\,Lallement and W. Reis for their help with the
improvements of the X-shooter response.
TR is
supported by the German Aerospace Center (DLR, grant 05\,OR\,1301.)
We thank an anonymous referee for helpful suggestions that
  improved this paper.
\end{acknowledgements}
\bibliography{Xshooter}
\end{document}